**Bernd Skiera, Lukas Jürgensmeier,**

**Kevin Stowe, Iryna Gurevych**

# How to Best Predict the Daily Number of New Infections of Covid-19

Bernd Skiera, Professor of Electronic Commerce and Member of the Board of EFL – The Data Science Institute, Goethe University Frankfurt, Theodor-W.-Adorno-Platz 4, 60629 Frankfurt, Germany, Phone +49-69-798-34649, email: skiera@wiwi.uni-frankfurt.de. Bernd Skiera is also (part-time) Professor Research Fellow at Deakin University, Australia.

Lukas Jürgensmeier, Member of the Leadership Team at TechAcademy e.V. and PhD Student at Graduate School of Economics, Finance and Management at Goethe University Frankfurt, 60629 Frankfurt, Germany, email: lukas.juergensmeier@tech-academy.io.

Kevin Stowe, PhD, Ubiquitous Processing Lab, Computer Science Department, Technical University of Darmstadt, Hochschulstrasse 10, 64289 Darmstadt, Germany, email: kevin.stowe@tu-darmstadt.de.

Iryna Gurevych, Full Professor, Ubiquitous Processing Lab, Computer Science Department, Technical University of Darmstadt, Hochschulstrasse 10, 64289 Darmstadt, Germany, email: iryna.gurevych@tu-darmstadt.de.

We acknowledge quick and very helpful feedback from Davide Difino, Oliver Hinz, Lennart Kraft, Maximilian Matthe, Celina Proffen, Vincent Skiera, and Cannon Warren.

# Abstract

Knowledge about the daily number of new infections of Covid-19 is important because it is the basis for political decisions resulting in lockdowns and urgent health care measures. We use Germany as an example to illustrate shortcomings of official numbers, which are, at least in Germany, disclosed only with several days of delay and severely underreported on weekends (more than 40%). These shortcomings outline an urgent need for alternative data sources. The other widely cited source provided by the Center for Systems Science and Engineering at Johns Hopkins University (JHU) also deviates for Germany on average by 79% from the official numbers. We argue that Google Search and Twitter data should complement official numbers. They predict even better than the original values from Johns Hopkins University and do so several days ahead. These two data sources could also be used in parts of the world where official numbers do not exist or are perceived to be unreliable.

# 1  Introduction

The daily number of new infections of Covid-19 is an important metric to track because it enables to determine how successful actions are in fighting the virus. These numbers are therefore the basis for political decisions resulting in lockdowns and urgent health care actions. Thus, accurate numbers are crucial to take the right actions and to save billions of Euro which may be lost due to incorrect decisions.

In Germany, the Robert Koch Institute (RKI) provides the official daily numbers of, among other data, new infections. Unfortunately, RKI does so with several days of delay and RKI also admits that their daily numbers on the weekend (i.e., Saturday and Sunday) are systematically lower than their daily numbers for weekdays (i.e., Monday to Friday) because the health departments of several individual states do not or only incompletely report cases on weekends (Merlot/Pauly 2020). Yet, RKI is silent about how much lower weekend numbers are.

It is obvious that a delay in reporting and inaccurate weekend numbers make decision making of politicians even more difficult than it already is. Not surprisingly, alternative data providers, in particular the Center for Systems Science and Engineering at Johns Hopkins University (JHU), filled the vacuum that RKI provided. Potential other alternative data providers, in particular the number of searches conducted on Google (subsequently referred to as Google Search) and the number of tweets on Twitter (referred to as Twitter data), were not considered despite their widespread availability. Daily search data on Google is disclosed about every week by Google and, while partly aggregated, extremely easy for everyone to observe. Daily Twitter data is more cumbersome to access but available without almost no delay.

Previous research has shown in different contexts that both data sources enable fairly accurate predictions in wide range of settings such as predicting stock prices, crime and the spread of flus (Achrekar et al. 2011, Du and Kamakura 2012, Hu, Du, and Damangir 2014, Gerber 2014, Joseph, Wintoki, and Zhang 2011, Nofer and Hinz 2015, Sakaki/Okazaki/Matsuo 2010). A particular advantage of these two data sources is that these (or related) data sources can potentially help even in situations in which official numbers are not available or unreliable, e.g., because of an inability to conduct medical tests or report test results. Such a situation might even occur in Germany and is likely to occur in other less developed parts of the world.

So far, however, it is not clear how good the prediction accuracy of these alternative data providers is. Thus, it is the aim of this manuscript to examine the ability of these alternative data sources, namely JHU, Google search data and Twitter data, to predict and complement the official numbers (as provided by RKI) of the daily number of new infections of Covid-19. Furthermore, we examine the underreporting of the official numbers over the weekend to get insights into the validity of the official numbers, indicating the need to complement these official numbers with other data sources such as the ones that we examine.

## 2  Information provided by Robert Koch Institute

The Robert Koch Institute (RKI) is the federal government entity in Germany managing disease control and prevention. In this capacity, it provides the data source of officially confirmed Covid-19 cases in Germany via API (Robert Koch Institute, 2020). It is based on information transmitted "daily" from the health departments of the 16 states in Germany. However, there are several problems with this data set. First, data is only consistently transmitted from the state health departments to RKI on working days. Data on weekends is therefore incomplete. Second, RKI often updates the numbers that were disclosed during the most recent days so that the "final" official numbers are only available a few days later (Merlot and Pauly 2020).

Unfortunately, RKI does not provide a detailed update history so that it is difficult to make a precise statement about this updating process. Figure 1 visualizes what we can derive from the publicly accessible data with respect to the size of RKI's late reporting on March 31, 2020 and April 5, 2020. It shows that the share of additional new infections that were reported on March 31 for March 29 is 66.2%. This value means that if on March 30 3,380 cases were reported for March 29, then the reporting on March 31 would update them to 10,000 cases. The share of additional new infections that were reported on March 31 for March 28 is 23.0%. Thus, the number of new infections is substantially underreported when they are initially disclosed. Looking at the right panel in the same figure, this pattern also holds in similar magnitudes for data disclosed on April 5, 2020. While the share of late registered cases is highest for the previous few days, RKI keeps adding and to a lesser degree subtracting confirmed Covid-19 cases from the official numbers until even one month into the past. JHU also changes earlier released numbers when new information for the previous days becomes available (JHU CSSE 2020).

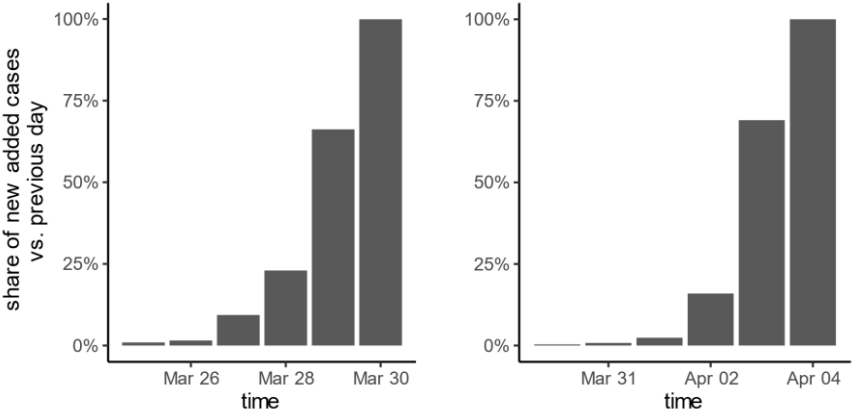

*Figure 1: Percentage of Newly Added Number on March 31 and April 5 of New Infections of Daily Total by Robert Koch Institute (RKI). Data Retrieved on March 31, 2020 (left panel) and April 5, 2020 (right panel).*

A problem that goes beyond the scope of this article is that the official number of new infections of Covid-19 only covers the number of new infections that are known, not the true number of new infections. This dark number is very likely to be much higher for a number of reasons. Among them are that people might be infected, but not display symptoms, people might not seek medical assistance or those who seek medical assistance might not be tested.

# 3 Alternative Data Sources

As RKI provides the official numbers late, we discuss three alternative data sources. The first is the data provided by Johns Hopkins University (JHU) that is widely used, and the two others are Google Search Data and Twitter data.

## 3.1 Information provided by Johns Hopkins University

The Center for Systems Science and Engineering (CSSE) at Johns Hopkins University (JHU) provides worldwide case numbers gathered from several different sources via their online dashboard and public GitHub repository (JHU CSSE, 2020). According to Dong et al. (2020), JHU derives the number of new infections by using official data sources from the World Health Organization, the American Center for Disease Control and the European Centre for Disease Prevention and Control but also less official data sources of new confirmed cases such as the information provided on Twitter, online news articles, and other information sources.

We start by comparing the daily number of new infections according to JHU with those from RKI. As outlined above, RKI updates its numbers even several days after its first reporting. Therefore, we only look at the number of new infections up to March 28, 2020 (i.e., seven days before the write-up of this manuscript) to make sure that we looked at the "final" official numbers. We started the comparison on March 3, 2020, when the daily number of new infections got larger than 50 in Germany.

Figure 2 in its upper panel visualizes the number of new infections per day and outlines the percentage difference in its lower panel. The mean absolute percentage error (MAPE) is 79.0% and JHU deviates in both directions: it overestimates (on weekends and on other days) but also underestimates. Note that March 7, 14, 21 and 28 are Saturdays, which occur two days before the Mondays that are displayed on the x-axes. JHU predicts higher numbers for these Saturdays and the Sundays.

*Figure 2: Comparison of Official Robert Koch Institute (RKI) and Predicted Johns Hopkins University (JHU) Daily Number of New Infections of Coronavirus Covid-19 in Germany (Data Retrieved April 5, 2020)*

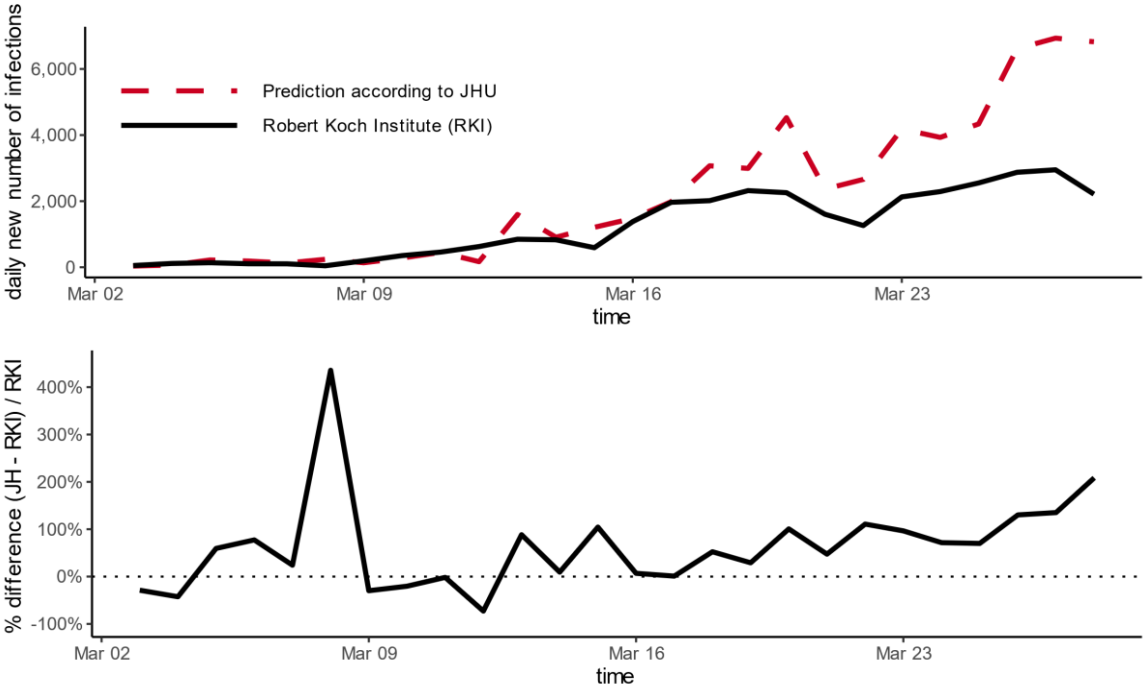

## 3.2  Google Search Data

As people increasingly turn to the Internet to gather news and information, their online search activity can be viewed as a snapshot of their collective consciousness, reflecting the interests, concerns and intentions of the population. Accordingly, Choi/Varian (2012) demonstrated that search volume data from Google accurately and timely tracks economic indicators (such as unemployment levels), whereas Goel et al. (2010) show that search volume data from Google predicts the population's future behaviour several days in advance. Within the public health context, Ginsberg et al. (2009) demonstrated that search volume for flu-related terms (such as "flu" and "cold") exhibits strong predictive power for the number of influenza cases in the same week (correlation of 0.94). Therefore, we examine if search volume data from Google could also be used to predict the number of Covid-19 infections – potentially several days in advance.

Through Google Trends (2020), we obtain publicly available daily search volume time series data for the term "corona". The data covers the time from January 19, 2020 – five days before the first confirmed case in Germany – up until March 28, 2020. To disguise the total number of searches, Google partly aggregates the data by indexing the maximum daily number of searches over the entire time span to a value of 100 and the values of all other days are calculated relatively to the maximum value. Figure 3 plots the time series of the relative search volume.

## 3.3 Twitter Data

Besides gathering information, people also use the Internet for spreading information themselves (for instance, on social media platforms such as Twitter). Similar to their search activity, such spreading of information can also potentially reveal people's current interests, concerns and intentions. In contrast to traditional media outlets, social media thereby has the major advantage that information can become available in almost real-time. Therefore, previous research has used Twitter data for prediction in various domains. For instance, Sakaki/Okazaki/Matsuo (2010) used Twitter data to detect earthquakes in Japan, Gerber (2014) used Twitter data for crime prediction in the US, and Nofer and Hinz (2015) used it for the prediction of stock prices. Within the public health context, Achrekar et al. (2011) and Paul et al. (2015) demonstrated that Twitter data improves the accuracy of flu prediction models. Therefore, we inspect how well Twitter data predicts the number of Covid-19 new infections.

*Figure 3: Indexed Number of Searches on Google and Tweets on Twitter Including the Term "Corona"*

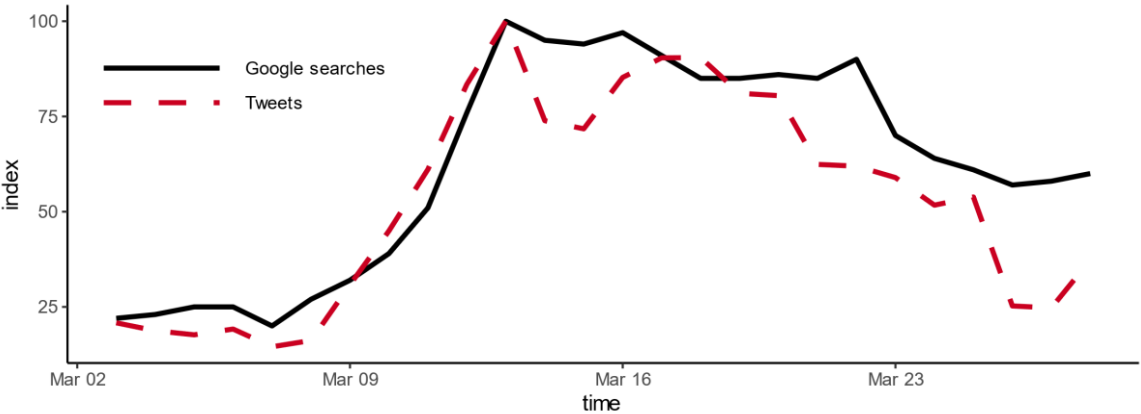

We collected data from Twitter via the Twitter advanced search interface, employing the NASTY web crawler (Schmelzeisen 2020), which interfaces with Twitter as a web browser allowing for powerful and efficient retrieval of tweets over time, to capture data from January 20, 2020 until March 28, 2020. We use the search term "corona" to filter down from the approximately 500 million tweets per day. Additionally, Twitter provides information about the tweet language. Since we want to predict the number of new infections in Germany, we restrict our analysis to tweets in German.

## 4  Prediction Quality of Alternative Data Sources

Subsequently, we compare the prediction quality of the alternative data sources and also consider the underreporting of RKI during weekends. As the underreporting during the weekend is likely to refer to a similar drop in percentages but not absolute numbers, we run a log-log-model with the logarithm of the daily number of new infections as the dependent variable and two independent variables, namely the logarithm of each of the respective numbers (i.e., daily number of new infections according to JHU, number of Google searches and number of Tweets) and a binary variable for the weekend (being 1 if the observation occurs on either Saturday or Sunday and 0 otherwise).

An advantage of data from Google search and Twitter is that both could be leading indicators. That means that search or Twitter behavior today might enable to predict the number of new infections tomorrow, the day after tomorrow or even days beyond the day after tomorrow. The reason could be that people might fear that they have Covid-19 and inform themselves on Google today but still wait a few days before they get medical assistance and therefore might be recorded as an official case of a new infection. Comparable arguments hold for Twitter. Appendix 7.2 describes our statistical procedure for determining how many days in advance Google search data and Twitter data can predict the number of new infections. Our result is that Google Search best predicts two days ahead of time and Twitter best predicts three days ahead of time. For consistency, and since the differences are small, we use a lag of three days for both data sources.

*Table 1: Regression Results with Official Daily Number of New Infections of Covid-19 in Germany as Dependent Variable and Three Alternative Data Sources and a Weekend Dummy as Independent Variables*

|  | Dependent variable: | | |
|---|---|---|---|
|  | Log. Number of new infections acc. to RKI | | |
|  | (1) | (2) | (3) |
| Log. number of new infections acc. to JHU | 0.799*** (0.059) | | |
| Log. number of Google searches (lag of three days) | | 1.954*** (0.201) | |
| Log. number of tweets (lag of three days) | | | 1.724*** (0.212) |
| Weekend | -0.482** (0.206) | -0.498* (0.275) | -0.685** (0.318) |
| Constant | 1.176*** (0.413) | -0.955 (0.794) | 0.232 (0.803) |
| Number of Observations | 26 | 26 | 26 |
| $R^2$ | 0.892 | 0.807 | 0.746 |
| Adjusted $R^2$ | 0.882 | 0.790 | 0.724 |
| Mean Absolute Error (MAE) | 0.440 | 0.462 | 0.495 |
| Mean Absolute Percentage Error (MAPE) on original data | 0.349 | 0.508 | 0.598 |
| Residual Std. Error (df = 23) | 0.465 | 0.621 | 0.713 |
| F Statistic (df = 2; 23) | 94.688*** | 48.003*** | 33.726*** |

*Note:* $p<0.1$; ***p<0.05;*** $p<0.01$
*RKI: Robert Koch Institute, JHU: Johns Hopkins University*

Table 1 presents the results of the three regression models. In the first one, we regress the (logarithmic) number of new infections of JHU on those of RKI. The mean absolute error (MAE) on the logarithmic data is 44.0%, which corresponds to a MAPE on the original values of 34.9%. This value is much lower than the MAPE of 79.0% when we just compared the original values and did not consider a weekend effect.

In the next two models, we use Google search data (model 2) with a lag of three days and the relative amount of Tweets including the term "corona" with a lag of three days (model 3). Their MAE on the logarithmic data are 46.2% and 49.5% and the respective MAPE on the original data are 50.8% and 59.8% and, thus, worse than JHU. Yet, their predictions are made three days earlier. So, earlier prediction comes at a cost but the earlier prediction might justify this cost.

Very consistently, all three regressions outline very sizable "weekend-effects" of RKI data, that is an underreporting of the number of daily new confirmed infections by -38.2% (=exp(-0.482)-1) in case of JHU (model 1), -39.2% (=exp(-0.498)-1) in case of Google search (model 2), and -49.6% (=exp(-0.685)-1) in case of Twitter (model 3).

To visually inspect the prediction quality of JHU as well as Google search and Twitter data, Figure 4 presents the time series of actual RKI and JHU daily new confirmed cases in Germany along with the models' predictions of RKI cases according to JHU and two alternative data sources: the lagged number of Google search queries and tweets. It outlines that all three alternative data sources had problems in predicting the most recent numbers. JHU overestimated and Google search and Twitter data underestimated at the end of the time series. Only the adjusted prediction of JHU (according to the regression display as model (1) in Table 1) predicted well.

*Figure 4: Comparison of Official Daily Number of New Infections of Covid-19 in Germany as reported by RKI with Predictions from Johns Hopkins University (JHU), Google Searches and Twitter Data*

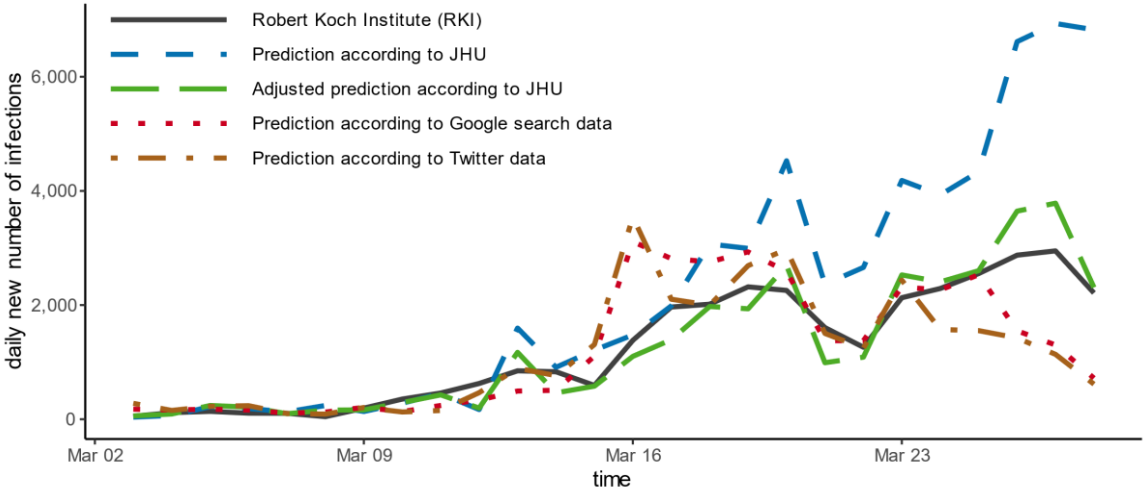

# 5  Summary, Conclusions and Recommendations

The daily number of new infections of Covid-19 is one of the most important metrics to detect whether actions towards fighting Covid-19 are successful. Unfortunately for Germany, the "final" official numbers are reported "correctly" with several days of delay, which means that the number of new infections on a particular day, say March 26, 2020, will be reported on the following day but updated again for several additional days, e.g., until March 30, 2020, or even later. This delay in reporting makes decision making even more difficult than it already is. It also makes the timely development of prediction models very challenging, particularly as Robert Koch Institute (RKI) does not provide a detailed update history of its disclosed numbers. An obvious but probably difficult to implement recommendation is to reduce the delay in reporting. A less obvious but important suggestion is to at least outline in detail the update history of the reported numbers of RKI.

A fairly consistent finding of the analysis with three alternative data sources is that the underreporting of RKI on weekends is between 38.2 and 49.6% depending on the model, thus in the area of more than 40%. We find it hard to justify such underreporting in times of great societal, humanitarian, and economic cost imposed by lockdowns. The economy loses billions of Euros because of actions that are based upon the information provided in the official reporting of daily new infections. We also wonder where those underreported infections on the weekend show up again. We admit that we did not analyze it in too much detail, but we could not identify a spike on Monday or Tuesday, which should have been the case if information that was collected on Saturday and Sunday would simply be reported one or two days later.

The rather large reporting error on weekends and the delay in publishing "final numbers" raise serious doubts about the validity of the official numbers in Germany and outline a need to complement the official numbers. The three alternative data sources that we examined have a prediction error (MAPE) between 34.9 and 59.5%. The adjusted numbers of Johns Hopkins

University (JHU) predict best. An advantage of Google search and Twitter data, however, is that they predict three days ahead of time. Thus, the predictions of all alternative data sources present an opportunity to complement the official numbers of RKI.

The difficult times that we face today encourage us to use our findings to elaborate on alternative data sources and additional questions that could be addressed. We focused on search data because Google invested into an infrastructure so that this data is easy to access for everyone (https://trends.google.com/). Twitter data is much harder to access but researchers, particularly in computer science have built up experience in mining this data. We did not use data from other popular websites such as Instagram or Facebook where consumers also express their preferences and needs. It would certainly help if the providers of such websites would at least make their "virus-related" data available and, if feasible while still protecting consumer privacy, would complement it with fair disaggregation at a regional level. Certainly, it would also be helpful if such data were available for other regions such as China, Iran or less developed countries in Africa or Asia, where other information sources might not exist or concerns about the validity of the official data are often expressed.

A challenging but important problem is also to make official data comparable across countries. Differences exist because of varying definitions of an infection and the amount of testing across countries. If valid, alternative data sources such as Google search data present the opportunity to at least complement the official data. However, such data is partly aggregated. Google search data, for example, report the relative popularity of search terms but not the precise search volume. The unknown precise transformation into relative popularity of terms, however, makes a comparison across countries almost impossible because volumes and proportions are likely to differ across countries. Therefore, we encourage providers such as Google to either disclose their "virus-related data" without partial aggregation, or to conduct their own analysis that uses original data to derive predictions across countries.

# 7  Online Appendix

## 7.1  Daily New Confirmed Covid-19 Cases in Germany According to RKI and JHU, and Relative Google Search Volume and Number of Tweets Containing the Term "Corona"

Table 2: *Daily Number of New Infections with Covid-19 in Germany According to Robert Koch Institute (RKI) and Johns Hopkins University (JHU), and Indexed Number of Searches on Google and Number of Tweets on Twitter Containing the Term "Corona" (Data Retrieved for RKI and JHU on April 5, 2020).*

| Date | Weekend | RKI | JHU | Google | Twitter |
| --- | --- | --- | --- | --- | --- |
| 2020-01-19 | TRUE | 0 | 0 | 0 | |
| 2020-01-20 | FALSE | 0 | 0 | 0 | 0 |
| 2020-01-21 | FALSE | 0 | 0 | 0 | 0 |
| 2020-01-22 | FALSE | 0 | 0 | 0 | 0 |
| 2020-01-23 | FALSE | 0 | 0 | 0 | 0 |
| 2020-01-24 | FALSE | 0 | 0 | 1 | 1 |
| 2020-01-25 | TRUE | 0 | 0 | 1 | 1 |
| 2020-01-26 | TRUE | 0 | 0 | 1 | 2 |
| 2020-01-27 | FALSE | 0 | 0 | 2 | 2 |
| 2020-01-28 | FALSE | 0 | 0 | 4 | 5 |
| 2020-01-29 | FALSE | 0 | 0 | 3 | 5 |
| 2020-01-30 | FALSE | 1 | 0 | 3 | 4 |
| 2020-01-31 | FALSE | 1 | 1 | 3 | 7 |
| 2020-02-01 | TRUE | 0 | 0 | 3 | 3 |
| 2020-02-02 | TRUE | 0 | 0 | 3 | 2 |
| 2020-02-03 | FALSE | 0 | 0 | 2 | 2 |
| 2020-02-04 | FALSE | 1 | 0 | 2 | 2 |
| 2020-02-05 | FALSE | 0 | 0 | 2 | 1 |
| 2020-02-06 | FALSE | 0 | 0 | 2 | 1 |
| 2020-02-07 | FALSE | 0 | 0 | 2 | 1 |
| 2020-02-08 | TRUE | 0 | 0 | 1 | 1 |
| 2020-02-09 | TRUE | 0 | 0 | 1 | 1 |
| 2020-02-10 | FALSE | 0 | 0 | 1 | 1 |
| 2020-02-11 | FALSE | 0 | 0 | 1 | 1 |
| 2020-02-12 | FALSE | 0 | 0 | 1 | 1 |
| 2020-02-13 | FALSE | 0 | 0 | 2 | 1 |
| 2020-02-14 | FALSE | 0 | 0 | 1 | 1 |
| 2020-02-15 | TRUE | 0 | 0 | 2 | 1 |
| 2020-02-16 | TRUE | 0 | 0 | 1 | 1 |
| 2020-02-17 | FALSE | 0 | 0 | 1 | 1 |
| 2020-02-18 | FALSE | 0 | 0 | 1 | 1 |
| 2020-02-19 | FALSE | 0 | 0 | 1 | 1 |
| 2020-02-20 | FALSE | 0 | 0 | 1 | 1 |
| 2020-02-21 | FALSE | 0 | 0 | 1 | 1 |
| 2020-02-22 | TRUE | 1 | 0 | 2 | 1 |
| 2020-02-23 | TRUE | 10 | 0 | 4 | 2 |
| 2020-02-24 | FALSE | 0 | 0 | 6 | 4 |
| 2020-02-25 | FALSE | 1 | 1 | 10 | 8 |
| 2020-02-26 | FALSE | 4 | 10 | 19 | 17 |
| 2020-02-27 | FALSE | 21 | 19 | 22 | 22 |
| 2020-02-28 | FALSE | 37 | 2 | 27 | 26 |
| 2020-02-29 | TRUE | 17 | 31 | 23 | 23 |
| 2020-03-01 | TRUE | 32 | 51 | 22 | 16 |

| Date | Weekend | RKI | JHU | Google | Twitter |
|---|---|---|---|---|---|
| 2020-03-02 | FALSE | 35 | 29 | 23 | 20 |
| 2020-03-03 | FALSE | 52 | 37 | 22 | 21 |
| 2020-03-04 | FALSE | 115 | 66 | 23 | 19 |
| 2020-03-05 | FALSE | 138 | 220 | 25 | 18 |
| 2020-03-06 | FALSE | 106 | 188 | 25 | 19 |
| 2020-03-07 | TRUE | 104 | 129 | 20 | 15 |
| 2020-03-08 | TRUE | 45 | 241 | 27 | 16 |
| 2020-03-09 | FALSE | 194 | 136 | 32 | 31 |
| 2020-03-10 | FALSE | 353 | 281 | 39 | 45 |
| 2020-03-11 | FALSE | 459 | 451 | 51 | 61 |
| 2020-03-12 | FALSE | 626 | 170 | 76 | 83 |
| 2020-03-13 | FALSE | 848 | 1,597 | 100 | 100 |
| 2020-03-14 | TRUE | 832 | 910 | 95 | 74 |
| 2020-03-15 | TRUE | 592 | 1,210 | 94 | 72 |
| 2020-03-16 | FALSE | 1,382 | 1,477 | 97 | 85 |
| 2020-03-17 | FALSE | 1,966 | 1,985 | 91 | 90 |
| 2020-03-18 | FALSE | 2,015 | 3,070 | 85 | 90 |
| 2020-03-19 | FALSE | 2,319 | 2,993 | 85 | 81 |
| 2020-03-20 | FALSE | 2,257 | 4,528 | 86 | 80 |
| 2020-03-21 | TRUE | 1,605 | 2,365 | 85 | 62 |
| 2020-03-22 | TRUE | 1,262 | 2,660 | 90 | 62 |
| 2020-03-23 | FALSE | 2,129 | 4,183 | 70 | 59 |
| 2020-03-24 | FALSE | 2,290 | 3,930 | 64 | 52 |
| 2020-03-25 | FALSE | 2,552 | 4,337 | 61 | 54 |
| 2020-03-26 | FALSE | 2,874 | 6,615 | 57 | 25 |
| 2020-03-27 | FALSE | 2,949 | 6,933 | 58 | 25 |
| 2020-03-28 | TRUE | 2,212 | 6,824 | 60 | 37 |
| 2020-03-29 | TRUE | 1,592 | 4,400 | | |
| 2020-03-30 | FALSE | 2,032 | 4,790 | | |
| 2020-03-31 | FALSE | 2,748 | 4,923 | | |
| 2020-04-01 | FALSE | 2,891 | 6,064 | | |
| 2020-04-02 | FALSE | 2,759 | 6,922 | | |
| 2020-04-03 | FALSE | 2,053 | 6,365 | | |
| 2020-04-04 | TRUE | 448 | 4,933 | | |

## 7.2 Determination of How Many Days Google Search and Twitter Data Predicts Ahead of Time

We investigate lags between zero and ten days and then use the adjusted R squared and the MAE to select the optimal lag for each of the two models. Figure 5 visualizes the prediction accuracy both in terms of adjusted R squared and MAE. We choose the third lag for Google Search as well as for Twitter data, because they approximately exhibit the highest adjusted R squared and lowest MAE at this lag of three days. Even though a lag of two days for the Google search model would yield slightly better results, we use a lag of three since the differences are small and enables us to be consistent with the Twitter model.

*Figure 5: Mean Absolute Error (MAE) and Adjusted R Squared of Regression Models with the Natural Logarithm of RKI's Daily Confirmed Cases as the Dependent Variable, one Lagged Logarithmic Value of Google Search (left) and Twitter (right) and a Weekend Dummy as the Independent Variable.*

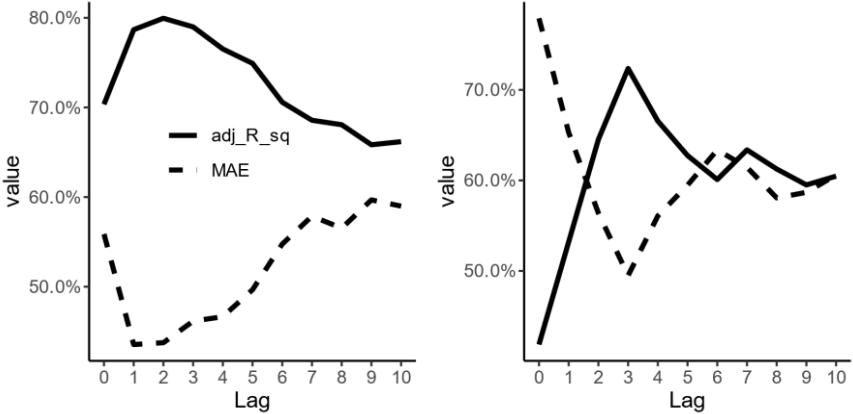